\documentstyle[12pt]{article}
\title{Neutrino mass and mirror universe}
\author{Z.~K.~Silagadze \vspace*{5mm}\\
\small\em JINR, Laboratory of Nuclear Problems, 141 980, Dubna, Russia
and \\ \small\em  Budker Institute of Nuclear Physics, 630 090,
Novosibirsk, Russia}
\date{}

\begin{document}
\large
\maketitle

\begin{abstract}
The existence of the mirror world, with the same
microphysics as our own one but with opposite {\bf P}-asymmetry,
not only restores an exact equivalence between left and right
but also naturally explains, via a see-saw like mechanism, why
the neutrino is ultralight.
\vspace*{5mm}
\newline  PACS number(s): 14.60.St, 14.60.Pq, 11.30.Er
\end{abstract}

 As is well known, {\bf P}-noninvariance of weak interactions
does not necessarily mean that there is an absolute difference between
left and right in nature. Any {\bf P}-asymmetry in matter (for example
the absence of right-handed neutrinos) can be accompanied by the opposite
{\bf P}-asymmetry in antimatter (for example the absence of
left-handed antineutrinos), so that the overall situation can still be
left-right symmetric. It is {\bf CP} (not {\bf P}) that represents
symmetry between left and right \cite{1}. Less formally, although
our world looks quite asymmetric when looked at through the {\bf P}-mirror,
it can still appear to be symmetric when looked at through the 
{\bf CP}-mirror.
But now we know that it is not either! In fact, it is a common
belief that left-right symmetry is associated with {\bf CP}, that
makes {\bf CP}-noninvariance so strange, otherwise a natural question
we should to ask is why {\bf CP}-violation is so tiny (and we must indeed
answer this question in QCD \cite{2}).

 However, in  contrast to this common belief, the left-right symmetry is not
necessarily associated with {\bf CP}-invariance. Space inversion (and
any other geometric symmetry from the Poincar\'e group) is represented
not by one quantum-mechanical operator {\bf P}, but by a whole class
of operators $\{ {\bf IP}\}$, where $\{ {\bf I}\}$ forms an internal
symmetry group of the system \cite{3}. Thus, we can use {\bf MP}, instead
of {\bf CP}, as a quantum-mechanical parity operator relating left and
right; here, {\bf M} is an {\bf arbitrary} internal symmetry operator. 
The question now arises of whether
we can find some sufficiently good internal symmetry for which {\bf MP} is
exactly conserved.

 In fact such an internal symmetry was proposed by Lee and Yang in
their seminal article \cite{4} about the possibility of parity
nonconservation, and it involves a drastic duplication of the world.
For any ordinary particle an existence of the corresponding "mirror"
particle is postulated, so that there are two kinds of electrons,
two kinds of photons and so on. The mirror world completely resembles
the ordinary one at the level of microphysics, except that it reveals
an opposite {\bf P}-asymmetry. In such an extended universe {\bf MP}
is an exact symmetry, where {\bf M} interchanges ordinary and mirror
particles, and there is no absolute difference between left and right:
this universe appears to be symmetric when looked at through the 
{\bf MP}-mirror.

Of course, mirror particles must interact with ordinary particles
very weakly to escape detection \cite{5}, but they must
interact at least gravitationally \cite{5}, so that sufficiently big 
clusters of mirror matter can cause observable gravitational effects
\cite{6}.
It is even possible that such effects have already been observed, if we
interpret dark matter as a mirror matter \cite{7}. 
The recent observation of the possible gravitational microlensing
events \cite{8} can then appear to be nothing but the observation of mirror
stars \cite{9}!

If there exist particles that carry both ordinary and mirror electric
charges (a connector), they can cause significant mixing
between ordinary and mirror photons even for a very heavy connector;
as a result mirror charged particles from the mirror world acquire
a small ordinary electric charge \cite{7,10}. Such millicharged
particles had been sought but never found \cite{11}. Another
consequence of the above photon mixing would be a
possibility for positronium to "disappear" in a vacuum (to oscillate
into a mirror positronium) \cite{12}. From available
experimental data on orthopositronium, it follows that photon-mirror 
photon mixing,
if any, is very small \cite{13} and that, most probably, the mixed form
of matter carrying both ordinary and mirror electric charges does not
exist \cite{12}.

 The ordinary and mirror universes can be grand-unified either with
a $G \times G$ type gauge group \cite{14} or even more tightly with
SO(n) type groups \cite{15}. In the latter case, the existence of such
queer objects as Alica strings is possible \cite{15,16}. An ordinary
particle encircling around this string transforms into a the mirror
particle. Thus, standard particles might go through the looking-glass
by means of such strings, as Alica did \cite{17}. This can lead to
observable astronomical effects. For example, if an Alica string
passes between the Earth and a galaxy, the galaxy becomes invisible
for a terrestrial observer \cite{16}.

 The most serious test for the scenario of the mirror world is provided
by cosmology \cite{18,19}, because the new degrees of freedom 
can effect the big bang nucleosynthesis \cite{20} and overproduce the
primordial ${^4}{\rm He}$. However, in contrast to the previous claims
\cite{18,19}, it appears that there are enough dodges for the mirror world
to pass this test \cite{9,21}.

 Rather unexpectedly, one more support for the mirror world hypothesis
comes from superstring theories. Namely, it was shown that some
heterotic string models lead, in the low energy limit, to the promising
$E_8 \times E_8$ effective gauge theory, with second $E_8$ acting in
the "shadow" world of mirror particles \cite{22}.

 A detailed analysis of observational physics of the mirror world and
a broad program of searches for astronomical effects of mirror matter
was given by M.Yu.~Khlopov et al. \cite{23}.

 Let us emphasize that, in order that the mirror world restore an exact 
left-right symmetry, 
it is sufficient that it exists in principle, with the same microphysics as 
in the ordinary world. Macrophysics can be quite
different. However, left-right symmetry is a rather abstract concept.
Can we point to some more material evidence in favor of the existence of the 
mirror world? We would like to indicate in this article that such 
evidence {\bf does} indeed exist. Even if the arguments presented in \cite{9}
failed and if the mirror world with different macrophysics proved empty, 
Francesco Sizzi's opinion, cited in \cite{18}, that such
a mirror universe, just like the moons of Jupiter discovered by
Galileo, "can have no influence on the Earth and therefore
would be useless and therefore does not exist", would be inapplicable
because, even if the mirror matter is diluted away by inflation, it
still leaves a very clear signal of its existence, and this is the
{\bf very small mass of the neutrino!}

 Actually, a possible connection between the mirror world and the neutrino
mass was already hinted in \cite{15} and \cite{24}. We are going to
argue that the assumption \cite{24} that small neutrino mass maybe
is a thin thread leading to the mirror world is indeed correct.

 Although many models have been proposed to explain the huge mass
difference between the neutrino and the corresponding charged lepton (see,
for example, \cite{25} and references therein), the most elegant
explanation is given by the so called seesaw mechanism \cite{26}. In
its original form, it leads to a naturally small Majorana mass for the
left-handed neutrino, but some modifications make it possible to
produce a small Dirac neutrino mass as well \cite{27,28}. 
The existence of other kinds of neutrinos, which are singlets under
the electroweak gauge group, is postulated in the later case. 
It should be noted, however, that the
universe with the mirror world is ideally suited for a seesaw-like
mechanism resembling that described in \cite{28}!

 For the sake of simplicity, we consider only one generation and assume that
the gauge group is $G_{WS}\times G_{WS}$ with conventional electroweak
group $G_{WS}=SU(3) \times SU(2) \times U(1)$. Further, we assume
that the known quarks and leptons together with their mirror partners,
transform as the $(f,1) \oplus (1,f^\prime)$ representation of $G_{WS}
\times G_{WS}$, where f is the ordinary quark-lepton family,
$$ f= {u \choose d}_L \hspace*{5mm} {\nu \choose e}_L
\hspace*{5mm} u_R \hspace*{5mm} d_R \hspace*{5mm} e_R \hspace*{5mm},$$
\noindent
and $f^\prime$ is that for mirror particles, except that left and
right are interchanged -- that is, $f^\prime$ contains right doublets and
left singlets with respect to the mirror weak isospin:
$$ f^\prime= {u^\prime \choose d^\prime}_R \hspace*{5mm} {\nu^\prime
\choose e^\prime}_R \hspace*{5mm} u^\prime_L \hspace*{5mm} d^\prime_L
\hspace*{5mm} e^\prime_L \hspace*{5mm}.$$
\noindent
In order that the neutrino not be discriminated, we also assume that 
there exist
a right-handed neutrino $\nu_R$ and its left-handed mirror partner
$\nu^\prime_L$, which are $G_{WS}\times G_{WS}$ singlets. Such
particles naturally arise if, for example,
$G_{WS}\times G_{WS}$ is a low energy remnant of $SO(10) \times
SO(10)$ grand unification.

 If there exists some scalar field $\varphi$, which is a singlet under
$G_{WS}\times G_{WS}$, the possible Yukawa coupling is
$\varphi ({\bar \nu}_R \nu^\prime_L+{\bar \nu}^\prime_L
\nu_R)$; if $\varphi$ develops a nonzero vacuum expectation value,
this will result in $\nu_R-\nu^\prime_L$ mixing $M({\bar \nu}_R
\nu^\prime_L+ {\bar \nu}^\prime_L \nu_R)$. Besides, ordinary
electroweak Higgs mechanism and its mirror partner will lead to
neutrino and mirror neutrino masses $m({\bar \nu}_L \nu_R+{\bar \nu}_R
\nu_L + {\bar \nu}^\prime_R \nu^\prime_L+{\bar \nu}^\prime_L
\nu^\prime_R)$, where m is expected to be on the order of the charged
lepton mass of the same generation. Note that {\bf MP} symmetry
guarantees the presence of only one mass parameter.

Nonzero $\langle \varphi \rangle $ does not affect $G_{WS}\times G_{WS}$ 
symmetry; therefore,
it is natural
to associate this vacuum expectation value with some early stages of
symmetry breaking in a more full theory (for example $SO(10) \times
SO(10) \to \linebreak SU(5) \times SU(5)$). Therefore, the expected
value of M is $10^{14}-10^{15}$\, Gev, and $\frac{m}{M}$ is indeed
very small.

 Thus, we expect the following neutrino mass terms
\begin{eqnarray} 
L_m =  M({\bar \nu}_R \nu^\prime_L+ {\bar \nu}^\prime_L
\nu_R)  + m({\bar \nu}_L \nu_R+{\bar \nu}_R \nu_L +
{\bar \nu}^\prime_R \nu^\prime_L+{\bar \nu}^\prime_L
\nu^\prime_R) 
\label{eq1} \end{eqnarray}

The mass eigenstates of (1) (physical neutrinos) are
\begin{eqnarray} &&
\tilde \nu_L = \cos \theta \, \nu_L+ \sin \theta \, \nu^\prime _L
\hspace*{10mm}
\tilde \nu^\prime_R = \cos \theta \, \nu^\prime_R+
\sin \theta \, \nu_R \equiv {\bf MP}(\tilde \nu_L)
\nonumber \\ && \vspace*{5mm}
\tilde \nu^\prime_L = \cos \theta \, \nu^\prime_L-
\sin \theta \, \nu _L \hspace*{10mm}
\tilde \nu_R = \cos \theta \, \nu_R- \sin \theta \, \nu^\prime _R
\equiv {\bf MP}(\tilde \nu^\prime _L) \hspace*{2mm}.
\label{eq2} \end{eqnarray}
Substituting (2) into (1), we immediately obtain
$$ \tan {(2\theta)}=-2r$$
\noindent
and
\begin{eqnarray}
L_m=M_+\left({\bar {\tilde \nu}}_R \tilde \nu^\prime_L+
{\bar {\tilde \nu}}^\prime _L \tilde \nu_R\right)+
M_-\left({\bar {\tilde \nu^\prime}}_R \tilde \nu_L+
{\bar {\tilde \nu}}_L \tilde \nu^\prime_R\right) 
\hspace*{2mm},
\label{eq3} \end{eqnarray}
\noindent
where
$$ r=\frac{m}{M}\, , \; M_+=\frac{M}{2}\left( 1+\sqrt{1+4r^2}\right)
\, , \; M_-=\frac{M}{2}\left( 1-\sqrt{1+4r^2}\right) \, .$$

 Thus, we have a superheavy Dirac neutrino $(\tilde \nu ^\prime _L,
\tilde \nu_R)$. Formulas (2) show that this is a rather bizarre object --
its
left-handed component inhabits mostly the mirror world, while right-handed
component prefers our ordinary world. Besides, we have an ultralight
Dirac neutrino $(\tilde \nu _L,\tilde \nu ^\prime _R)$. Its left-handed
component $\tilde \nu _L$ is nearly our
old nice neutrino from $\beta$-decay, and the right-handed component
$\tilde \nu ^\prime _R $
is probably more familiar to mirror physicists.

 To summarize, the mirror-world hypothesis of Lee and Yang is very
attractive. Not only does this hypothesis restore full equivalence between 
left and right, but it can also explain why the neutrino has
an incredibly small mass.

 It should be noted that the idea of this study emerged while
reading the book \cite{29}.

After this work had been completed, I became aware of some relevant
articles. The possibility of exact parity conservation was rediscovered
by R.~Foot, H.~Lew and R.R.~Volkas in \cite{30}. The possible
consequences of this -- in particular, for neutrino physics -- were 
thoroughly investigated in \cite{31}. The effect of
the mirror world on neutrino physics was also considered in the recent
study by Z.~Berezhiani and R.~Mohapatra \cite{32}.

\section*{Acknowledgments}
 I am grateful to V. De Alfaro for information about massive
compact halo objects.

\end{document}